\documentclass[11pt, times,twocolumn]{article}

\usepackage{graphicx}
\usepackage{subfig}
\usepackage{amssymb}
\usepackage{cite}

\begin{document}
\topmargin 0pt
\oddsidemargin 0mm
\renewcommand{\thefootnote}{\fnsymbol{footnote}}
\begin{titlepage}

\vspace{5mm}

\begin{center}
{\Large \bf Strain-tunable band parameters of ZnO monolayer in graphene-like honeycomb structure} \\

\vspace{6mm} {\large Harihar Behera\footnote{E-mail: harihar@phy.iitb.ac.in;
 harihar@iopb.res.in } and Gautam Mukhopadhyay\footnote{Corresponding author's E-mail: gmukh@phy.iitb.ac.in; g.mukhopa@gmail.com}}    \\
\vspace{5mm}
{\em
Department of Physics, Indian Institute of Technology, Powai, Mumbai-400076, India} \\

\end{center}

\vspace{5mm}
\centerline{\bf {Abstract}}
\vspace{5mm}
  We present ab initio calculations which show that the direct-band-gap,
effective masses and Fermi velocities of charge carriers in ZnO monolayer
(ML-ZnO) in graphene-like honeycomb structure are all tunable by application
of in-plane homogeneous biaxial strain. Within our simulated strain limit of
$\pm 10\%$, the band gap remains direct and shows a strong non-linear
variation with strain. Moreover, the average Fermi velocity of electrons
in unstrained ML-ZnO is of the same order of magnitude as that in graphene.
The results promise potential applications of ML-ZnO in mechatronics/straintronics
and other nano-devices such as the nano-electromechanical systems (NEMS) and
nano-optomechanical systems (NOMS).\\

\end{titlepage}

\section{Introduction}
Nanomaterials whose electronic properties can be controlled by mechanical
   strain are highly desirable for applications in nano-electromechanical
   systems (NEMS). If such materials happen to have direct band gaps, then
   those can find applications in nano-optomechanical systems (NOMS).
   If mechanical strain strongly affects the electronic properties of a
   material, then that material can be used as a strain sensor. Bulk ZnO,
   in its most stable wurtzite structure at ambient pressure, has many useful
   properties \cite{1,2} such as direct and wide band gap (E$_{\mbox g} = 3.37$
   eV at room temperature and 3.44 eV at low temperatures), large exciton
   binding energy (60 meV), strong piezoelectric and pyroelectric properties,
   strong luminescence in the green-white region, large non-linear optical
   behavior, high thermal conductivity, radiation hardness, biocompatibility,
   and so on. These properties make ZnO an excellent candidate for a variety of
   applications in optics, electronics and photonics, sensors, transducers and
   actuators. Several ZnO nanostructures in the form of very thin nanosheets,
   nanowires, nanotubes and nanobelts have been synthesized and characterized
   using different preparation methods \cite{3,4}. In 2006, using the density
   functional theory (DFT) calculations, Freeman et al. \cite{5} first predicted
   that when the layer number of ZnO films with (0001) orientation is small,
   the wurtzite structures are less stable than a phase based on 2D ZnO sheets
   with a layer ordering akin to that of hexagonal BN. In 2007, Tusche et al.
   \cite{6} reported the observation of 2 monolayer (ML) thick ZnO(0001) films
   grown on Ag(111) by using surface X-ray diffraction and scanning tunneling
   microscopy. Very recently graphene-like honeycomb structures of ZnO have
   been prepared \cite{7} on Pd(111) substrate. Theoretical studies
   \cite{5,8,9,10} have now established the stability of the monolayer and
   few-layers (FL) of ZnO in planar honeycomb structures. This stability
   is attributed to the strong in-plane sp$^2$ hybridized bonds between the
   Zn and O atoms.\\
\indent
   Recent theoretical studies on two dimensional (2D) flat ZnO nanostructures
   reported (i) the elastic, piezoelectric, electronic, and optical properties
 of ZnO monolayer (ML-ZnO)\cite{10}, (ii) room-temperature half-metallic
 ferromagnetism in the half-fluorinated ML-ZnO \cite{11}, (iii)
 fluorination induced half metallicity in few ZnO layers \cite{12}, (iv)
 fully-fluorinated and semi-fluorinated ZnO sheets \cite{13} and (v)
 strain-induced semiconducting-metallic transition for ZnO zigzag nanoribbons
 \cite{14}. Here, we report our DFT based investigation of the effect of
 homogeneous in-plane biaxial strain on the electronic properties of ML-ZnO -
 a specific study not found in the existing literature. This study simulates
 an experimental situation where ML-ZnO is supported on an ideal flat
 stretchable/flexible substrate.


\section{Computational Methods}
 The calculations have been performed by using the DFT based full-potential
   (linearized) augmented plane wave plus local orbital (FP-(L)APW+lo) method
   \cite{15} which is a descendant of FP-LAPW method \cite{16}. We use
   the elk-code \cite{17} and the Perdew-Zunger variant of LDA \cite{18},
   the accuracy of which have been successfully tested in our previous works
   on graphene and silicene \cite{19}(silicon analog of graphene), germanene \cite{20}
   (germanium analog of graphene)and graphene/h-BN heterobilayer \cite{21}.
   The plane wave cut-off of $|{\bf G} +{\bf k}|_{max}= 9.0/R_{mt}$ (a.u.$^{-1}$) ($R_{mt}=$ the smallest muffin-tin radius) was chosen for plane wave expansion in the interstitial region. The Monkhorst-Pack \cite{22} ${\bf k}$-point grid size of $30\times30\times1$ was used for all calculations. The total energy was converged within $2\mu$eV/atom. We simulate the 2D-hexagonal structure of ML-ZnO as a 3D-hexagonal supercell with a large value of $c$-parameter ($= |{\bf c}| = 40$ a.u.). The application of in-plane homogeneous biaxial strain $\delta$ up to $\pm 10\%$ was simulated by varying the in-plane lattice parameter $a (=|{\bf a}| = |{\bf b}|); \delta = (a - a_0)/a_0$, where $a_0$ is the ground state in-plane lattice constant. Figure 1 depicts the top-down view of a ML-ZnO in planar configuration.
\begin{figure}[h]
 \centering
 \includegraphics[scale=0.75]{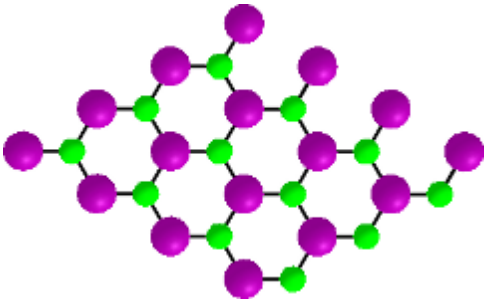}
 \caption{Top-down view of a ZnO monolayer in graphene-like honeycomb structure
  in ball-stick model. The large (pink) ball represents Zn atom and the small
  (green) ball represents the O atom.}
\end{figure}
\section{Results and Discussions}
\begin{figure}
 \centering
 \includegraphics[scale=0.85]{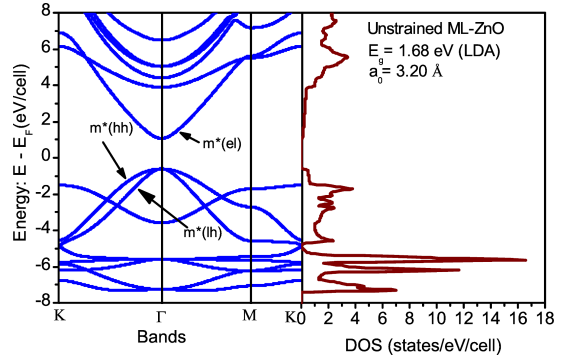}
 \caption{Bands and total DOS of unstrained ML-ZnO within LDA.
 $\mbox{E}_F$ is the Fermi energy.}
\end{figure}
 Our calculated LDA value of $a_0 = 3.20$ \AA \, corresponds to the Zn-O bond
length $d_{Zn-O} = a_0/\sqrt{3} = 1.848$ \AA,\, which is in agreement with the reported  experimental \cite{6} value of $d_{Zn-O} = 1.92$ \AA\, and theoretical
values of $d_{Zn-O} = 1.86$ \AA\, by Tu and Hu \cite{8}, 1.895 \AA\, by Topsakal
et al. \cite{9}, 1.853 \AA\, by Tu \cite{10}, 1.85 \AA\, by Wang \cite{12}. Our
$3.75\%$ underestimation of $d_{Zn-O}$ value with respect to the experimental
value is due to the well known problem of underestimation of the lattice
constant within LDA. In our previous studies \cite{18,19} of graphene,
silicene and germanene, we have demonstrated that the value of $c$-parameter
chosen in the construction of supercell for simulation of 2D hexagonal
structures also affects the value of in-plane lattice constant: larger value of
$c$-parameter yields a slight smaller value of $a$. Since we use a different
method and a different value of $c$-parameter, the disagreement of our result
on $a_0$ with other theoretical result \cite{9} is acceptable.\\
 \begin{figure}
 \centering
 \includegraphics[scale=0.85]{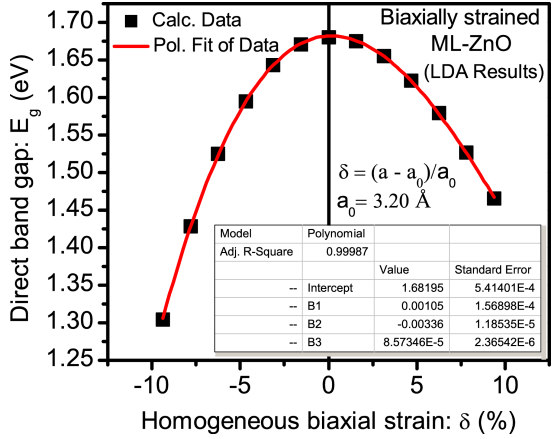}
 \caption{Variation of the direct band gap of ML-ZnO with in-plane homogeneous biaxial
 strain $\delta$. }
\end{figure}
\indent
  The electronic band structure and total density of states (DOS) plots of
  ML-ZnO are depicted in Figure 2. As seen in Figure 2, ML-ZnO is a direct
  band gap (E$_g$ = 1.68 eV, LDA value) semiconductor with both valence band
  maximum (VBM) and conduction band minimum located at the $\Gamma$ point of the
hexagonal Brillouin Zone (BZ). However, the actual band gap is expected to be
larger as LDA is known to underestimate the gap. Our calculated LDA band gap of
1.68 eV is in agreement with previous calculations of 1.68 eV by Topsakal et al.
 \cite{9}, 1.762 eV by Tu \cite{10}, 1.84 eV by Wang et al. \cite{13}. Using
the GW approximations, recently Tu \cite{10} estimated the direct band gap of
 ML-ZnO at 3.576 eV, which showed ML-ZnO as a wide band gap semiconductor.
 Although both LDA and generalized gradient approximations (GGA) do not yield
 the band gap correctly, these are powerful enough to predict the correct trend in variation
 of the gap \cite{23,24,25,26}. Since we focus on the trend as well as the nature of
  variation in band gap (such as the possible transition from direct to indirect
  gap-phase as reported theoretically for uniaxially strained ZnO nanotubes \cite{27}
  and ZnO nanowires and nanotubes \cite{28}) rather than its absolute value,
  we employed the computationally simpler and less time-consuming LDA for
  investigating the effect of biaxial strain on band parameters of ML-ZnO. 
 \begin{figure}
 \centering
 \includegraphics[scale=0.90]{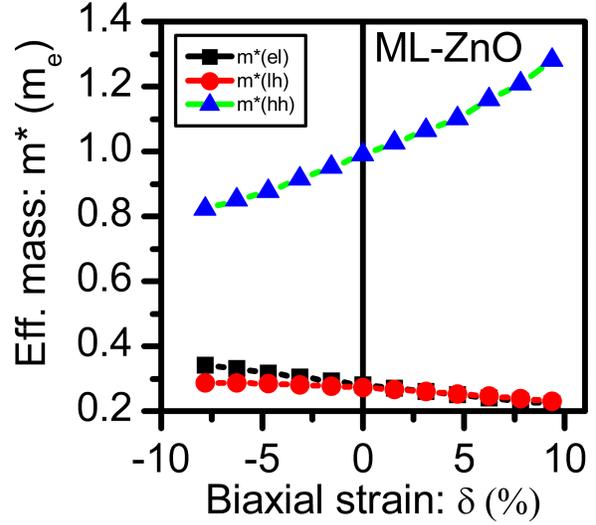}
 \caption{Variation of effective masses of the electron $m^\star$(el),
 light hole $m^\star$(lh) and heavy hole $m^\star$(hh) at the $\Gamma$ point
 of the BZ with in-plane homogeneous biaxial strain.}
\end{figure}
\begin{figure}
 \centering
 \includegraphics[scale=0.90]{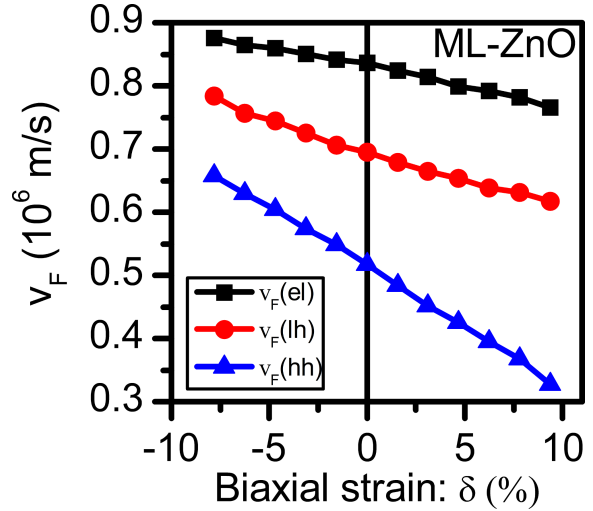}
 \caption{Variation of Fermi velocities of the electron $v_F$(el),
 light hole $v_F$(lh) and heavy hole $v_F$(hh) at the $\Gamma$ point of the
  BZ with in-plane homogeneous biaxial strain.}
\end{figure}
\indent
     Our calculated results on homogeneous biaxial strain-induced modifications
 of the band gap of ML-ZnO are depicted in Figure 3, which shows a strong
 non-linear variation of band gap with biaxial strain.
 Within our simulated  strain limits 
 $(\approx -10\% < \delta <10\%)$ as shown in Figure 3, the band gap $\mbox{E}_g$
 remains direct and its value varies with $\delta \,(= [(a-a_0)/a_0)]\times 100\, \%)$ as
 \begin{equation}
 \mbox{E}_{g}(\delta)= B_0+ B_1\times \delta + B_2\times \delta ^2 + B_3\times \delta ^3
\end{equation}
where $B_0 = 1.68195$ eV, $B_1= 0.00105$ eV, $B_2 = - 0.00336$ eV,
$B_3 = 8.57346\times10^{-5}$ eV, obtained by polynomial fit of our computed data. This
may be useful in calculating the LDA band gap corresponding to a particular value of biaxial
strain $\delta$ in the given range.
Our prediction of strain-engineered band gap of ML-ZnO, if
 verified by experiments, may be useful in fabrication of NEMS and NOMS based
 on a graphene-analogue of ZnO. The energy bands very close to the $\Gamma-$point
 are parabolic, which enabled us to define and estimate the effective masses. Figure 4
 depicts the nature of variation of
 effective masses of electrons $m^\star$(el), light holes $m^\star$(lh) and
 heavy holes $m^\star$(hh) at the $\Gamma$-point of the BZ (see Figure 2) with
 our simulated biaxial stain. Slightly away from the parabolic bands close to
 the $\Gamma-$point, the bands are linear in $k$ which allowed us
to calculate the Fermi velocity $(v_F)$ values from their slopes.
 The nature of variation of our calculated Fermi velocities of electrons $v_F$(el), 
 light holes $v_F$(lh) and the heavy holes $v_F$(hh) near the $\Gamma$-point of 
 the BZ with our simulated biaxial stain
 is depicted in Figure 5. We found that the Fermi velocity $v_F$(el) of unstrained
 ML-ZnO ($v_F$(el) $= 0.836\times 10^6$ m/s for $\delta = 0$ in Figure 5) near the
 $\Gamma$-point of BZ is close to he experimentally observed value of
 $v_F = 0.79\times10^6$ m/s in monolayer graphene (MLG) deposited on graphite
 substrate \cite{29} and the theoretically calculated DFT value of
 $v_F = 0.833\times10^6$ m/s in MLG \cite{30} and $v_F \simeq 0.8\times10^6$
 m/s in MLG supported on monolayer of hexagonal boron nitride \cite{21,31}. Since
 the charge carriers enter into ballistic transport limit when they have high
 velocities in the absence of scattering \cite{32}, we expect ballistic transport
 in ML-ZnO due to the high velocity of charge carriers analogous to the case
 of graphene.
\section{Conclusions}
 Our all-electron full-potential density functional theory based calculations
 show that band-gap, effective masses and Fermi velocities of charge carriers
 in ML-ZnO in graphene-like honeycomb structure are mechanically tunable by
 application of in-plane homogeneous biaxial strain. Our predictions on strong
 non-linear variation of direct band gap with biaxial strain (Figure 3),
  non-linear increase of $m^\star$(el), and almost linear decrease in $m^\star$(lh) and $m^\star$(hh) with increasing strain (Figure 4) and almost linear decrease in the Fermi
  velocities of charge carriers with increasing strain (Figure 5) are not only
  novel but also experimentally testable by application of biaxial strain to a
  flat stretchable/flexible substrate supporting ML-ZnO. Our results are
  significant in that an analogous honeycomb-structure `graphene' does not
  show any variations of its band gap under similar biaxial strains and currently
  a lot research is being done not only to open but also to control the band gap in
  graphene. Owing to its direct band gap with strong non-linear variation with
  biaxial strain, ML-ZnO should have potential applications in mechatronics/straintronics
   and other nano-devices such as the NEMS and NOMS.

\bibliographystyle{elsarticle-num}

\end{document}